\documentclass[twocolumn]{aastex6}
\pdfoutput=1 
\usepackage[utf8]{inputenc}
\usepackage{CJK}
\usepackage{amsmath,amstext}
\usepackage[T1]{fontenc}
\usepackage[figure,figure*]{hypcap}
\usepackage{affils}
\usepackage{hyperref}
\usepackage{microtype}
\usepackage{caption}
\usepackage{subcaption}
\usepackage{xspace}

\def\kms{km\,s$^{-1}$}

\newcommand{\cntext}[1]{\begin{CJK}{UTF8}{bsmi}#1\end{CJK}}

\shorttitle{Host galaxy of SLSN~2017egm}
\shortauthors{Chen et al.}

\begin{document}
\title{Spatially resolved MaNGA observations of the host galaxy of superluminous supernova~2017egm}
\DeclareAffil{mpe}{Max-Planck-Institut f{\"u}r Extraterrestrische Physik, Giessenbachstra\ss e 1, 85748, Garching, Germany; \href{mailto:jchen@mpe.mpg.de}{jchen@mpe.mpg.de}}
\DeclareAffil{nz}{Department of Physics, University of Auckland, Private Bag 92019, Auckland, New Zealand}
\DeclareAffil{subaru}{Subaru Telescope National Astronomical Observatory of Japan, 650 N Aohoku Pl., Hilo, HI 96720, USA}
\DeclareAffil{IANCU}{Graduate Institute of Astronomy, National Central University, No. 300, Zhongda Rd., Zhongli Dist., Taoyuan City 32001, Taiwan}
\DeclareAffil{qub}{Astrophysics Research Centre, School of Mathematics and Physics, Queen's University Belfast, Belfast BT7 1NN, UK}
\DeclareAffil{sh}{Department of Physics and Astronomy, University of Southampton, Southampton, SO17 1BJ, UK}

\affilauthorlist{Ting-Wan Chen \cntext{(陳婷琬)}\affils{mpe},
Patricia Schady\affils{mpe}, Lin Xiao\affils{nz}, J.J. Eldridge\affils{nz}, Tassilo Schweyer\affils{mpe}, \\
Chien-Hsiu Lee \cntext{(李見修)}\affils{subaru}, Po-Chieh Yu \cntext{(俞伯傑)}\affils{IANCU}, Stephen J. Smartt\affils{qub} and Cosimo Inserra\affils{sh}
}

\begin{abstract}
Superluminous supernovae (SLSNe) are found predominantly in dwarf galaxies, indicating that their progenitors have a low metallicity. However, the most nearby SLSN to date, SN~2017egm, occurred in the spiral galaxy NGC~3191, which has a relatively high stellar mass and correspondingly high metallicity. In this paper, we present detailed analysis of the nearby environment of SN~2017egm using MaNGA IFU data, which provides spectral data on kiloparsec scales. 
From the velocity map we find no evidence that SN~2017egm occurred within some intervening satellite galaxy, and at the SN position most metallicity diagnostics yield a solar and above solar metallicity ($12+\log{\rm(O/H)} \sim 8.8-9.1$).
Additionally we measure a small \mbox{H$\alpha$}\, equivalent width (EW) at the SN position of just 34\,\AA, which is one of the lowest EWs measured at any SLSN or Gamma-Ray Burst position, and indicative of the progenitor star being comparatively old. 
We also compare the observed properties of NGC~3191 with other SLSN host galaxies. 
The solar-metallicity environment at the position of SN~2017egm presents a challenge to our theoretical understanding, and our spatially resolved spectral analysis provides further constraints on the progenitors of SLSNe. 
\end{abstract}


\section{Introduction}
\label{sec:intro}

Superluminous supernovae (SLSNe)\footnote{In this paper we use the term SLSN to refer only to hydrogen-poor, Type I SLSNe} have been discovered in recent wide field, untargeted surveys \citep[e.g.][]{2011Natur.474..487Q}. 
They are 100 times brighter \citep[absolute magnitude of $-21$;][]{2012Sci...337..927G} than typical core-collapse SNe, and the standard paradigm of iron-core collapse cannot account for their origin. 
Further constraints on the progenitor properties arise from studying the nearby environments. 
SLSNe appear to exclusively occur in dwarf galaxies \citep{2011ApJ...727...15N, 2013ApJ...763L..28C, 2014ApJ...787..138L, 2015MNRAS.449..917L, 2016MNRAS.458...84A} with less than around a half-solar metallicity \citep{2016ApJ...830...13P, 2017MNRAS.470.3566C, 2016arXiv161205978S}. Their host galaxies also typically have high specific star formation rates (sSFR$\equiv$SFR/$M_{*}$), which may indicate that the progenitors are very young stars \citep{2015MNRAS.449..917L}.
Given that low mass and metal-poor galaxies have typically larger sSFR, it is difficult to disentangle the relative importance of these properties towards the formation of SLSNe \citep{2016ApJ...830...13P,2017MNRAS.470.3566C}. From the ejecta masses \citep{2015MNRAS.452.3869N}, and oxygen masses in the 
nebular spectra \citep{2017ApJ...835...13J}, it seems certain that the progenitors are above 20\mbox{M$_{\odot}$\,}, but whether they exclusively arise 
from higher mass stars ($M>40$\mbox{M$_{\odot}$\,}) remains to be seen.

\begin{figure*}
    \centering
    \begin{subfigure}[h]{0.40\linewidth}
        \includegraphics[width=\columnwidth]{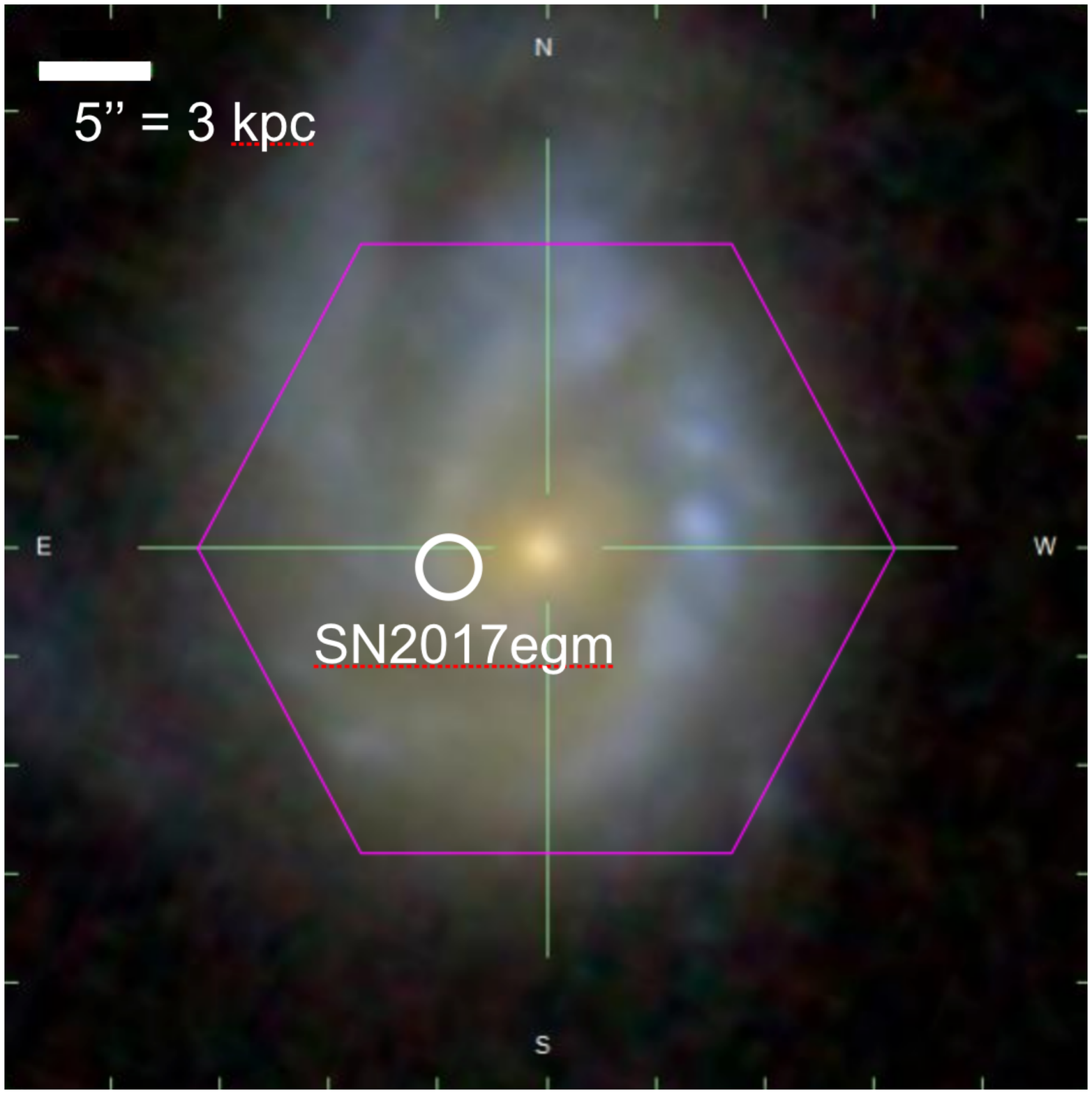}
    \end{subfigure}  
~~
    \begin{subfigure}[h]{0.45\linewidth}
        \includegraphics[width=\columnwidth, trim={1.5cm 0cm 2.5cm 0.25cm}]{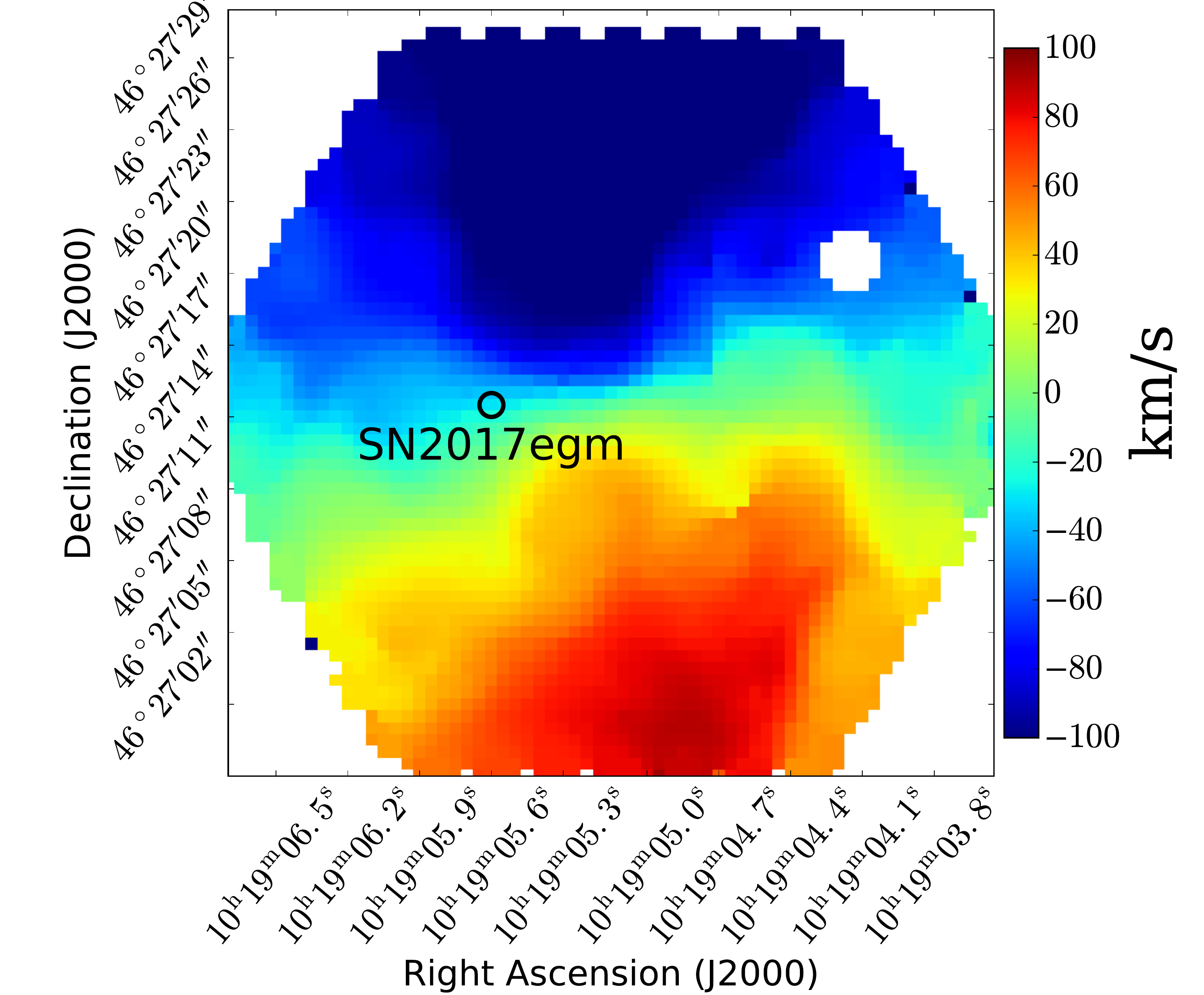}
    \end{subfigure}
\caption{{\it Left:} MaNGA FOV of NGC~3191. 
{\it Right:} Velocity map inferred from the position of the H$\alpha$ line relative a redshift of $z=0.031$. In the top right region of the map there are about 30 pixels that show an additional feature just redward of the H$\alpha$ line that is clearly not intrinsic to the nebular line emission from NGC~3191, we therefore masked-out the area where this unidentified feature appears.}
\label{fig_fov_velocity}
\end{figure*}

The recent discovery of the closest SLSN to date, SN~2017egm at $z=0.0307$, which occurred in the massive, spiral galaxy NGC~3191 challenges the current hypothesis on the required host properties (\citealt{2017ATel.10498...1D, 2017arXiv170608517N} and \citealt{2017arXiv170800864B}).   
One major issue with these previous studies is they do not probe the local environment of SN~2017egm. The two available SDSS
spectra analysed in \cite{2017arXiv170608517N} were from spectrograph fibers that did not cover the SN site.
Here we present a detailed analysis of the host galaxy of SN~2017egm using SDSS/MaNGA\footnote{\texttt{http://www.sdss.org/surveys/manga}} survey \citep[Mapping Nearby Galaxies at APO;][]{Bundy+15} data obtained from the SDSS Data Release 14\footnote{\texttt{http://skyserver.sdss.org/dr14/en/home.aspx}}. The observation was conducted on 2016-Jan-17.
The MaNGA field-of-view provides spectral measurements across a large area of NGC~3191 (Fig. \ref{fig_fov_velocity}).
Using these data, we present the direct measurement of metallicity at this SLSN site, use the EW to study the progenitor properties, and compare the gas-phase properties of NGC~3191 with other SLSN host galaxies.

\begin{figure*}[t!] 

\begin{subfigure}{0.48\textwidth}
\includegraphics[width=\linewidth]{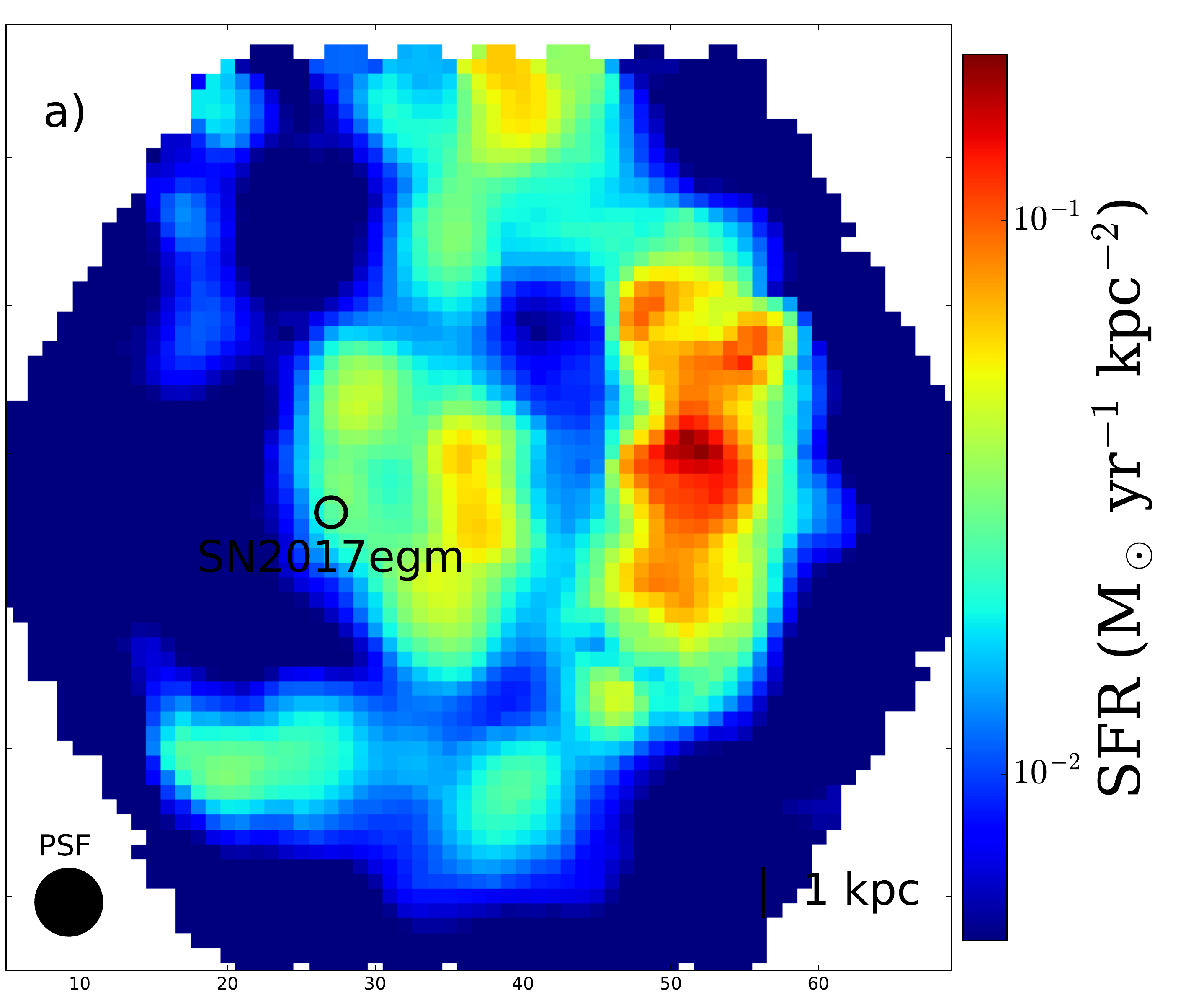}
\end{subfigure}\hspace*{\fill}
\begin{subfigure}{0.48\textwidth}
\includegraphics[width=\linewidth]{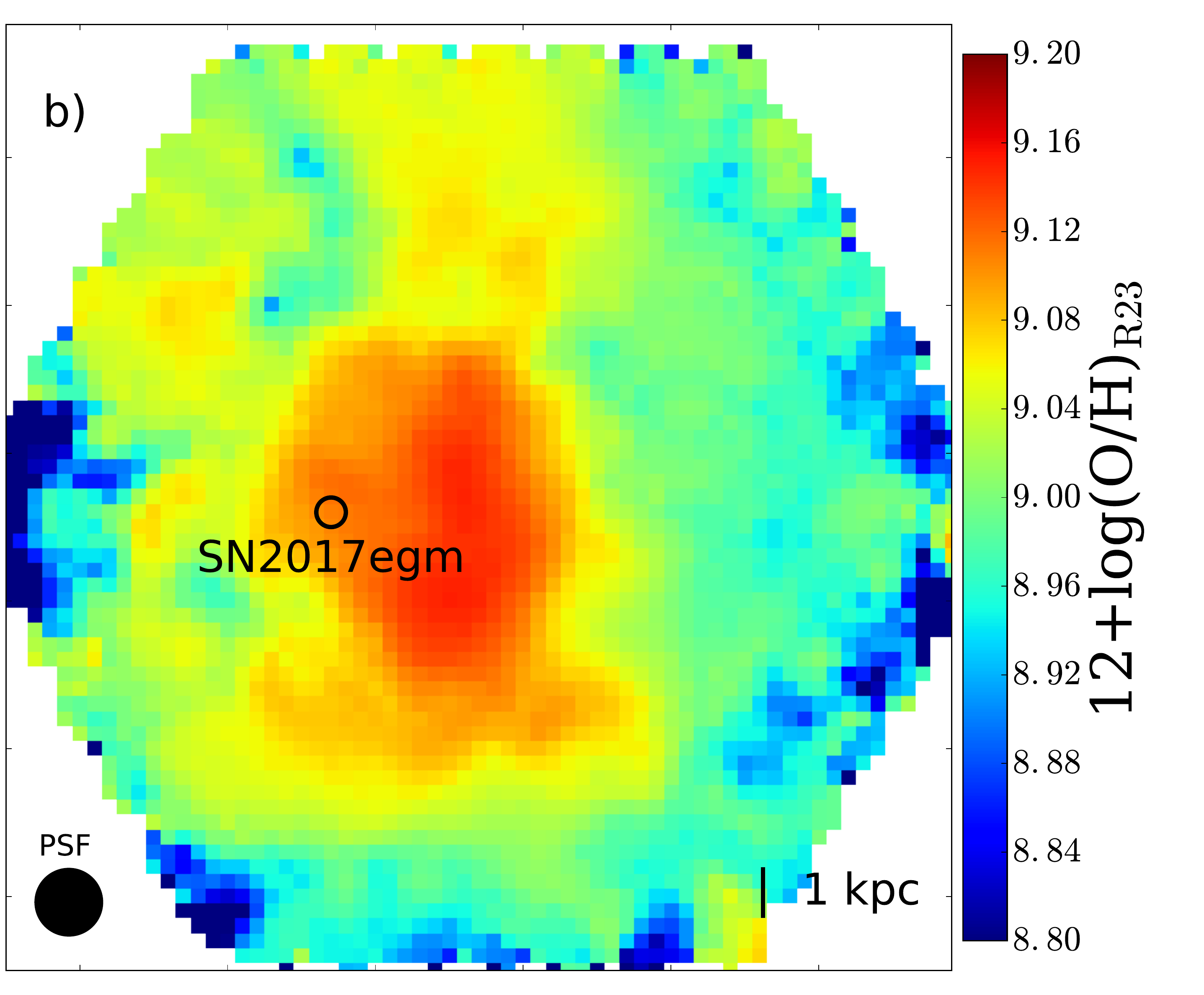}
\end{subfigure}

\medskip
\begin{subfigure}{0.48\textwidth}
\includegraphics[width=\linewidth]{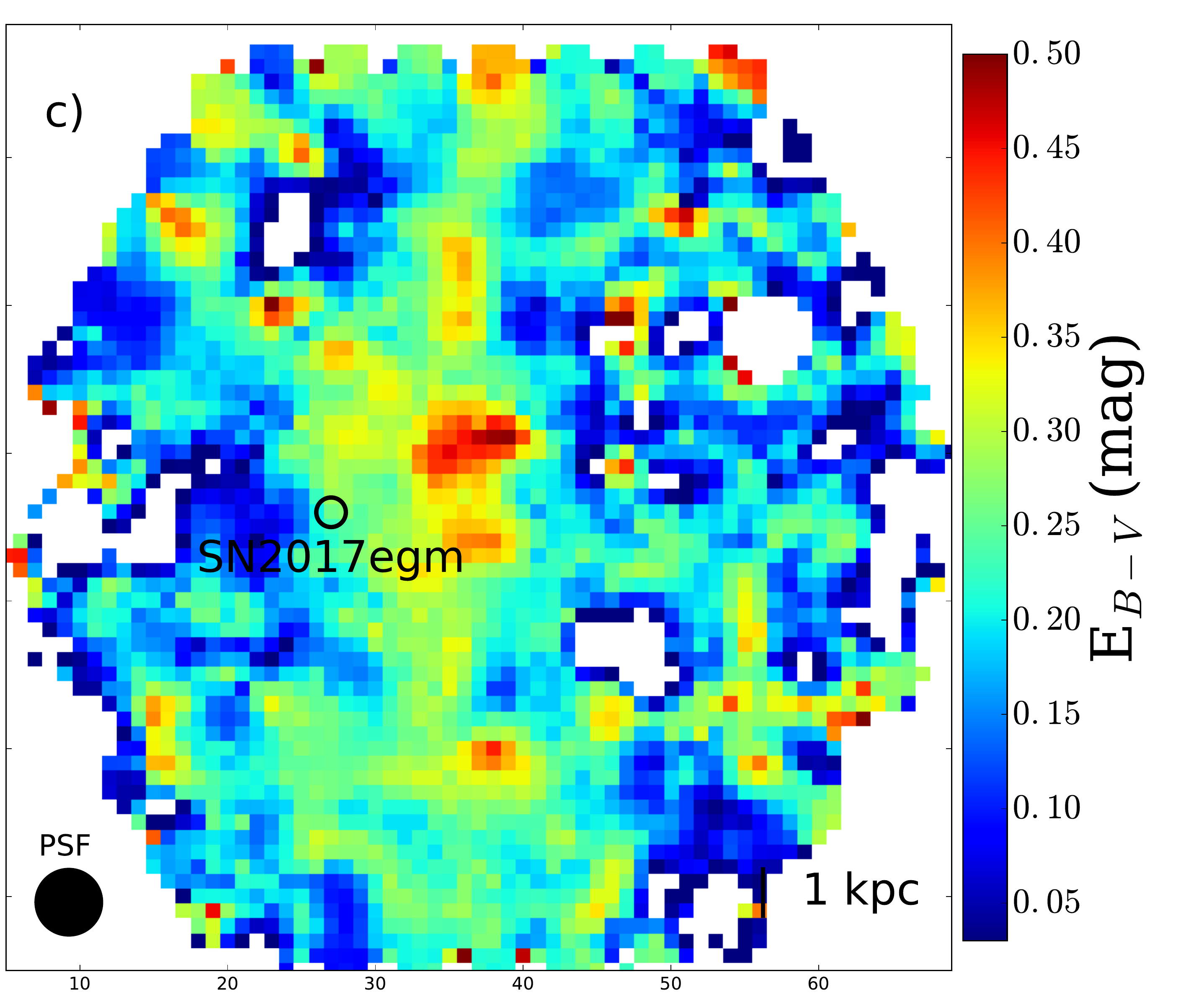}
\end{subfigure}\hspace*{\fill}
\begin{subfigure}{0.48\textwidth}
\includegraphics[width=\linewidth]{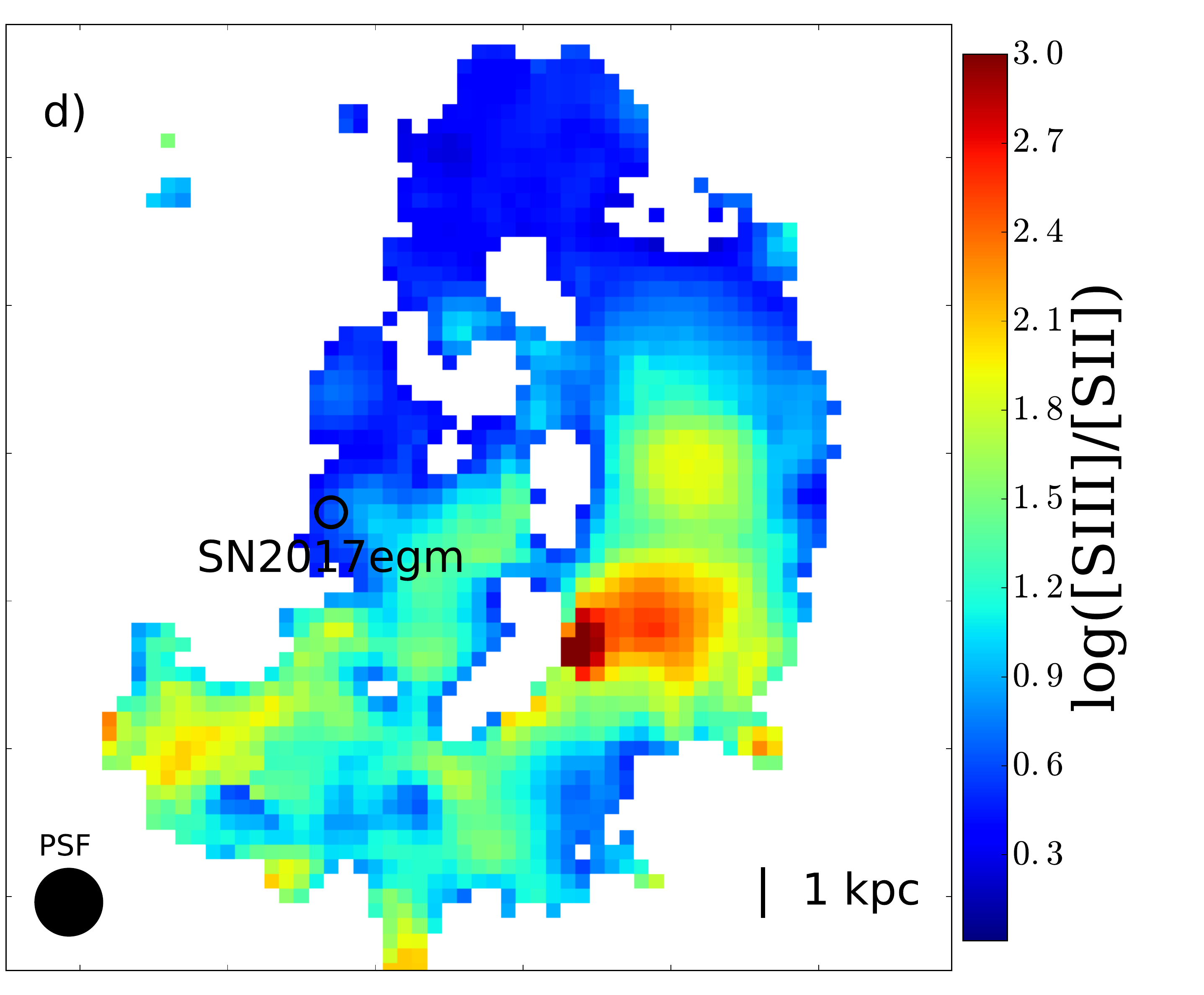}
\end{subfigure}

\medskip
\begin{subfigure}{0.48\textwidth}
\includegraphics[width=\linewidth]{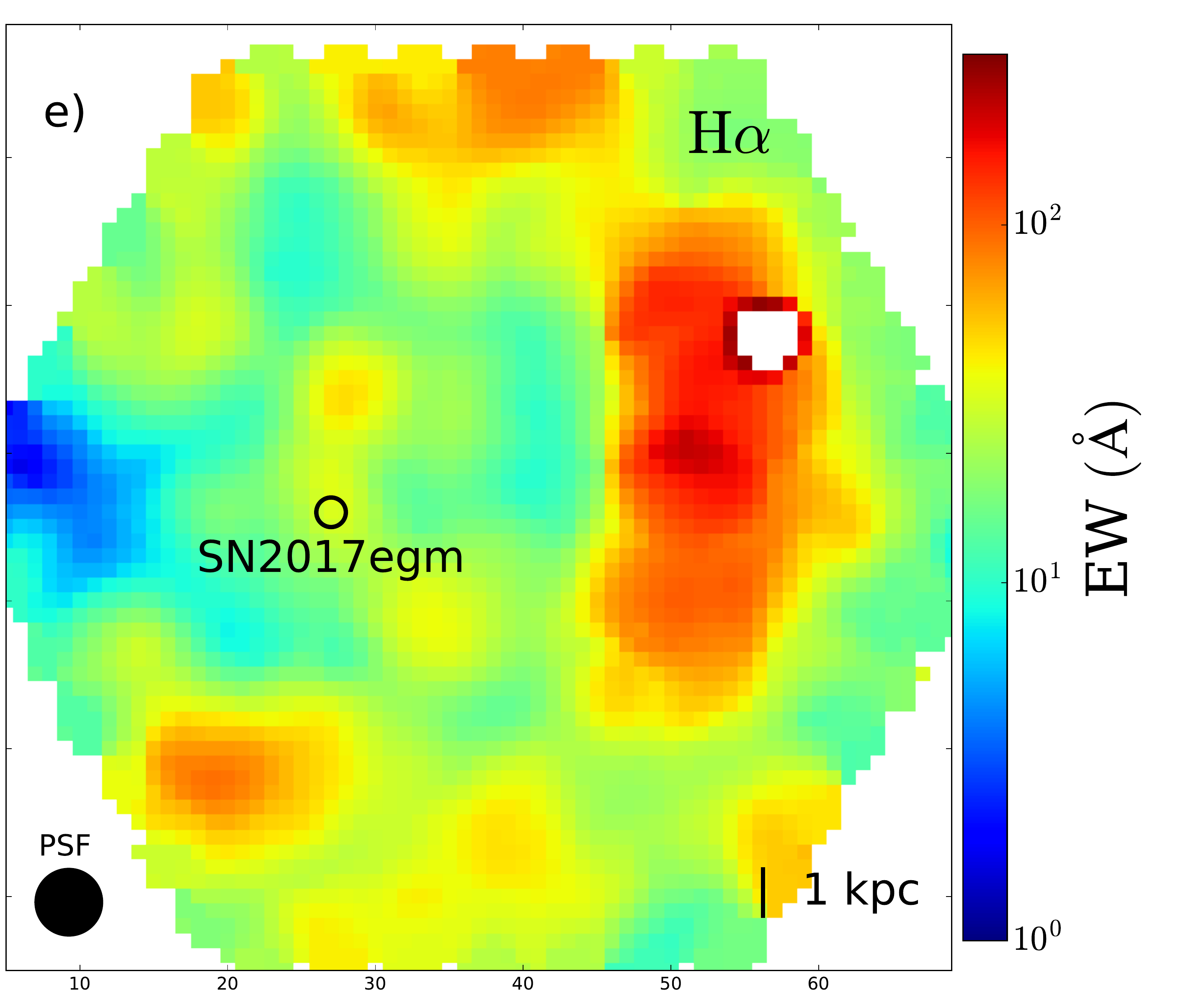}
\end{subfigure}\hspace*{\fill}
\begin{subfigure}{0.48\textwidth}
\includegraphics[width=\linewidth]{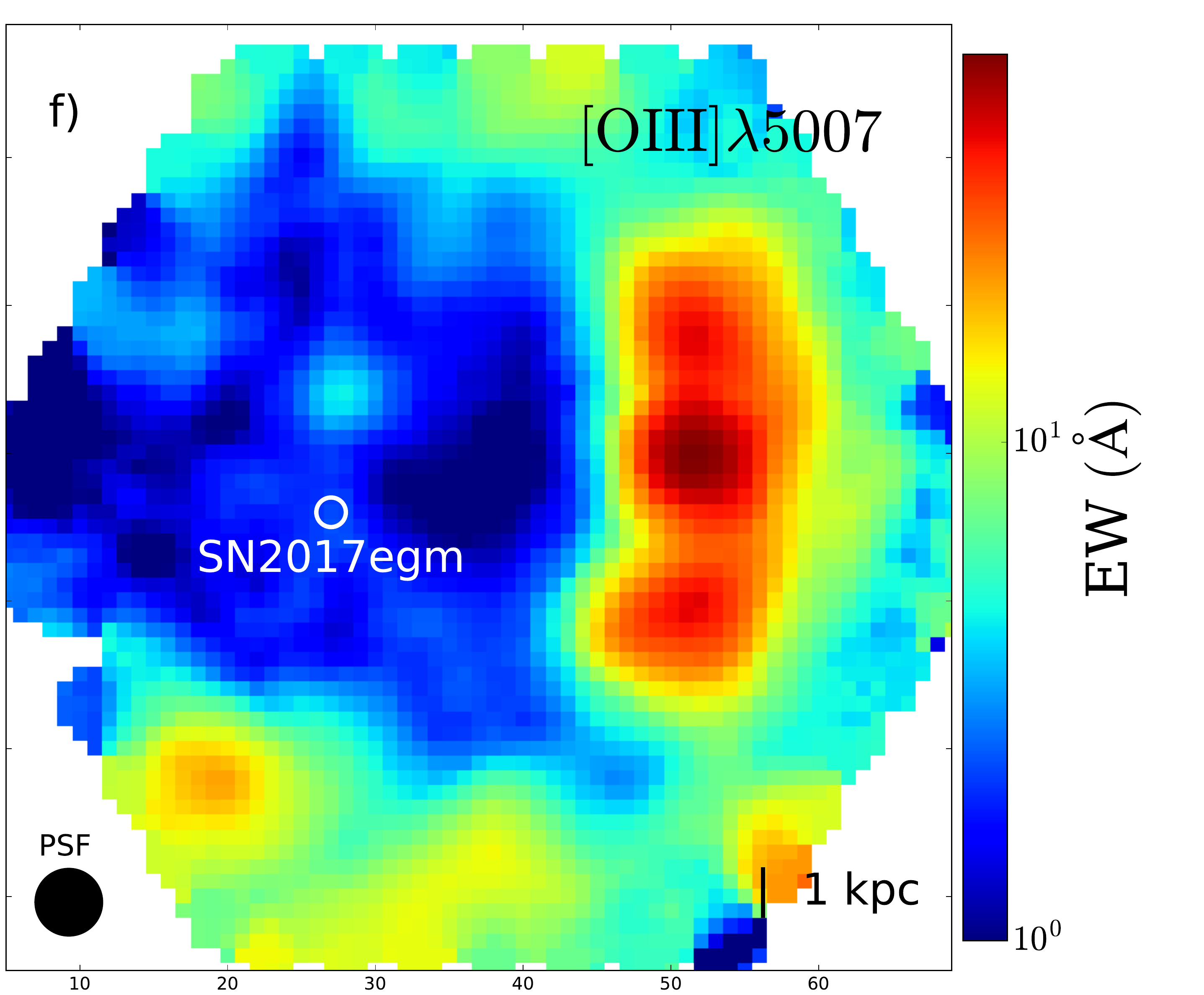}
\end{subfigure}

\caption{Reconstructed images of NGC~3191 from the MaNGA data cube.
{\it (a):} SFR map. 
{\it (b):} Metallicity map in KK04 $R_{23}$ scale.
{\it (c):} Dust-reddening distribution based on the H$\alpha$/H$\beta$ Balmer decrement
{\it (d):} The [\ion{S}{3}]/[\ion{S}{2}] flux ratio, which is a proxy for ionization. 
In c) and d) only spaxels with S/N $>3$ are shown.
{\it (e):} H$\alpha$ EW map.
{\it (f):} [\ion{O}{3}] EW map.
The images are approximately $30\times 35$\,arcsec, corresponding to a physical size of about $19\times 21$\,kpc at the redshift of NGC~3191. The effective spatial resolution is given by the PSF indicated in the lower left of each figure, which has FWHM of approximately 2.6\,arcsec, or 1.6\,kpc, and the physical scale is plotted in the bottom right corner.}
\label{fig:all_map}
\end{figure*}

\begin{figure*}
    \centering
       \includegraphics[width=\columnwidth]{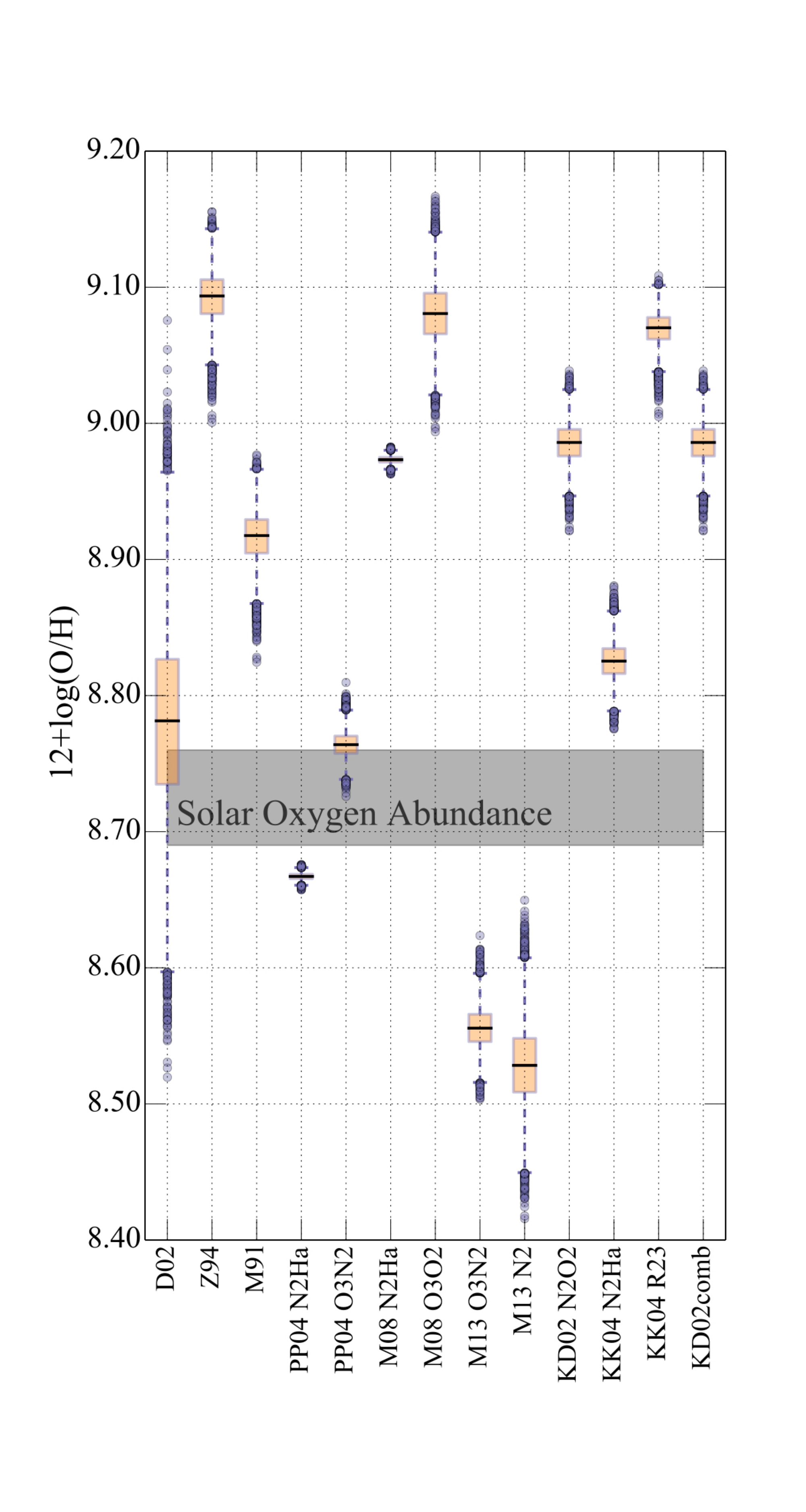}
\caption{Results of different metallicity diagnostics at the SN location using {\sc pymcz} \citep{2016A&C....16...54B}. {\sc pymcz} calculates the metallicity probability distribution for a given set of line fluxes with associated errors. For each metallicity diagnostic shown the black horizontal line corresponds to the medium value of the probability distribution, and the interquartile range is indicated by the orange box. The blue dashed lines represent the minimum and maximum of the distribution excluding outliers. The range in solar oxygen abundances reported in the literature is indicated by the gray region. }
\label{fig_Z}
\end{figure*}

\section{Results of the MaNGA data of NGC~3191}
\label{sec:results}
We used our own inhouse tools to analyze the large MaNGA datacube, following a similar procedure as described in \citet{2017A&A...602A..85K}.  
To separate the stellar and gas-phase components of the galaxy, we used the spectral synthesis code {\sc starlight} \citep[e.g.][]{2009RMxAC..35..127C} to fit stellar population models
to each spaxel. By subtracting the best-fit stellar template from the observed spectra, we were left with a datacube only containing the contribution of the gas-phase, which we then used to derive the velocity and metallicity maps.

\subsection{Velocity map}
Many spiral galaxies are often accompanied by dwarf satellite galaxies (e.g. the Magellanic satellites of the Milky Way), and thus SN~2017egm might reside in a nearby dwarf satellite galaxy along the line-of-sight, rather than in NGC~3191 itself. 
Such a kinematically distinct galaxy (200-300\,\kms might be expected) could be observed in 
emission lines if the SFR of the satellite were high enough. 

We created a velocity map by measuring the position of the \mbox{H$\alpha$}\, line with respect to the rest-frame wavelength at each spatial pixel, assuming a redshift of $z=0.0307$. 
As shown in Fig.\,\ref{fig_fov_velocity}, the velocity map is smooth across the SN~2017gem explosion site with no indication of 
a kinematically distinct component. Using the \mbox{H$\alpha$}\, flux at the SN position to derive our sensitivity to the detection of a foreground dwarf galaxy, we place a conservative upper limit of $f_{{\rm H}\alpha}<1.9\times10^{-16}$\,erg\,s$^{-1}$\,cm$^{-2}$ (which corresponds to SFR $< 0.002$\,\mbox{M$_{\odot}$\,}yr$^{-1}$).
We compared this limit with a nearby ($z<0.3$) SLSN host sample \citep{2017MNRAS.470.3566C}, and only 2 of 19 hosts may have a SFR less than our limit (the remaining 17 hosts have SFRs $>0.01$\,\mbox{M$_{\odot}$\,}yr$^{-1}$). Therefore, if a line-of-sight satellite would be the true host galaxy, it is at the lowest end of the host SFR distribution. On the other hand, the satellite would have to be many times larger than the spatial resolution of our MaNGA data to have a comparative SFR to the mean SFR in the \citealt{2017MNRAS.470.3566C} sample.

\subsection{Metallicity map}
Using pre-explosion MaNGA data gives us a unique opportunity to investigate the local metallicity at the SN~2017egm explosion site and to study the possibility that low-metallicity regions exist within the galaxy. 

After correcting for foreground reddening and internal dust extinction using the Balmer decrement (for example, $E(B-V) = 0.26\pm0.04$\,mag at the SN location, see Fig.\,\ref{fig:all_map}),
we measured main galaxy emission line fluxes in each spaxel. 
We applied the `$R_{23}$' strong line diagnostic with the \citet{2004ApJ...617..240K} (thereafter KK04) scale, which has the advantage that it fits for ionization parameter and metallicity iteratively. 
We used the [\ion{N}{2}]/[\ion{O}{2}] to distinguish between the two $R_{23}$ solutions \citep{2008ApJ...681.1183K}. At the SN position $\log$([\ion{N}{2}]/[\ion{O}{2}])=$-0.32$, corresponding to the upper branch solution of $12+\log {\rm(O/H)} = 9.11\pm0.01$, which is equivalent to 2.6\,\mbox{Z$_{\odot}$\,} assuming a solar oxygen abundance of 8.69 \citep{2009ARA&A..47..481A}.

From the metallicity map, we see no evidence of low-metallicity gas clumps around the SN region.
We also measured the metallicity using the O3N2 diagnostic \citep[][thereafter PP04]{2004MNRAS.348L..59P}, which gave a value of $12+\log {\rm(O/H)} = 8.77\pm0.01$ (1.3\,\mbox{Z$_{\odot}$\,}) at the SN position. 
The O3N2 map and $R_{23}$ maps are very similar, and thus we only provide $R_{23}$ in Fig.\,\ref{fig:all_map}.
Furthermore, there is no apparent correlation between the $R_{23}$ metallicity and the ionization maps (Fig.\,\ref{fig:all_map}) as traced by [\ion{S}{3}]/[\ion{S}{2}] \citep{1991MNRAS.253..245D}, as expected \citep{2017A&A...602A..85K}.

The metallicity diagnostics are known to be uncertain \citep{2008ApJ...681.1183K}, so we used the open-source python code {\sc pymcz} \citep{2016A&C....16...54B} to calculate the oxygen abundance in several strong line diagnostics. Figure.~\ref{fig_Z} shows the range in metallicities 
although most diagnostics give a metallicity at the location of SN~2017egm that is above solar. 
One of the few exceptions are the M13 diagnostics \citep{2013A&A...559A.114M}, which were the diagnostics considered in \citet{2017arXiv170803856I}. We note that the metallicity that we measured using the M13 O3N2 diagnostic is consistent with the values reported in \citet{2017arXiv170803856I}, and thus differences in the metallicities between the two papers is a result of only the metallicity diagnostic, and not in the measured line fluxes.

\subsection{EW map}
In Fig.\,\ref{fig:all_map} we show the \mbox{H$\alpha$}\, and [\ion{O}{3}] EW maps of NGC~3191. 
SN~2017egm is located in a star-forming region, and at the SN position we measured an \mbox{H$\alpha$}\, EW of $33.7\pm3.3$\,\AA. This is far lower than the \mbox{H$\alpha$}\, EW in the most active star-forming region of the galaxy (Fig.\,\ref{fig:all_map}), on the west spiral arm, also clearly seen as several bright \ion{H}{2} regions in the SDSS images.  

We measured an EW of [\ion{O}{3}] equal to $2.2\pm0.2$\,\AA\ at the SN site, which is surprisingly low compared to what is measured in other spatially resolved SLSN environments. Extreme emission line EWs
have been suggested to be a defining property of SLSN host galaxies, and indicative that SLSNe arise from the youngest of stars \citep{2015MNRAS.449..917L}. 
Strong [\ion{O}{3}] is also indicative of low metallicity (due
to lack of cooling through oxygen ions and subsequently high $T_{\rm e}$). The relatively low [\ion{O}{3}] EW supports the region being both metal rich and is not
indicative of a very young stellar population.
The [\ion{O}{3}] EW in the most star forming region of the galaxy is $> 100$\,\AA. 
If NGC~3191 had been at a higher redshift (and spatially unresolved), we may have incorrectly concluded that SN~2017egm also exploded within a region of very young star formation. At the spatial resolution of the MaNGA data, the emission at the SN location likely comes from several \ion{H}{2} regions or an OB complex, and thus the EW values quoted above are an average.

The total EW of [\ion{O}{3}] integrated over the entire MaNGA data cube is 8.0\,\AA\ and the total EW of \mbox{H$\alpha$}\, is 42.7\,\AA.
From \citealt{2015MNRAS.449..917L}, the median EW of [\ion{O}{3}] of 13 SLSN host galaxies is 190\,\AA, and the median EW of \mbox{H$\alpha$} of 12 SLSN host galaxies is 177\,\AA. Therefore, the EW of NGC~3191 is at the lowest end of the EW distribution of SLSN host galaxies.

Additionally, we reconstructed the SFR map (Fig.\,\ref{fig:all_map}) from the \mbox{H$\alpha$} flux, assuming a Chabrier IMF. The local SFR at the SN position is about 0.04\,\mbox{M$_{\odot}$\,}yr$^{-1}$, while the total SFR is 4.4\,\mbox{M$_{\odot}$\,}yr$^{-1}$.

\section{Discussion and conclusions}

\subsection{Outliers, the most metal-rich host galaxy of SLSNe}
There are approximately 90 SLSNe which have been identified to date\footnote{For a complete SLSN list see the website: \texttt{https://slsn.info}, and references therein}, around 60 of which have a measured host stellar mass (median of $\log(M_{*}/{\rm \mbox{M$_{\odot}$\,}})=8.3$) and around 30 of them have a measured gas-phase metallicity (median of 0.3\,\mbox{Z$_{\odot}$\,} using the PP04 O3N2 scale). 

In the nearby Universe ($z<0.3$), PTF~10uhf \citep{2016ApJ...830...13P} is a clear outlier, with the highest stellar mass ever measured for a SLSN host galaxy of $\log(M_{*}/{\rm \mbox{M$_{\odot}$\,}})=11.2$. Although the SN is far away from the host centre  (17.5\,kpc), the metallicity in the $R_{23}$ KK04 scale at the SN site (8.8\,dex) is similar to that at the nucleus (9.0\,dex).
SN~2017egm has a smaller offset of 3.1\,kpc from the host centre, and it has a host stellar mass of $\log(M_{*}/{\rm \mbox{M$_{\odot}$\,}})=10.3-10.7$.
From the MaNGA data we measured a metallicity of $12+\log {\rm(O/H)} = 9.11\pm0.01$ (2.6\,\mbox{Z$_{\odot}$\,}, in the $R_{23}$ KK04 scale) at the SN site, which is the most metal-rich environment for any SLSN discovered so far. This is in conflict with the 0.5\,\mbox{Z$_{\odot}$\,} threshold suggested for the formation of SLSNe \citep{2016ApJ...830...13P, 2017MNRAS.470.3566C} and challenges a pair-instability SN (PISN) as the energy source of at least SN~2017egm, with PISN being unlikely to be formed at \mbox{Z$_{\odot}$\,}$>0.2$ \citep{2013MNRAS.433.1114Y}.  We note that [O/H] does not relate well to [Fe/H], and so by saying it is oxygen rich, does not necessarily mean that it is iron rich, although the two are related.

\begin{figure*}
    \centering
    \begin{subfigure}[h]{0.45\linewidth}
    \includegraphics[width=\columnwidth]{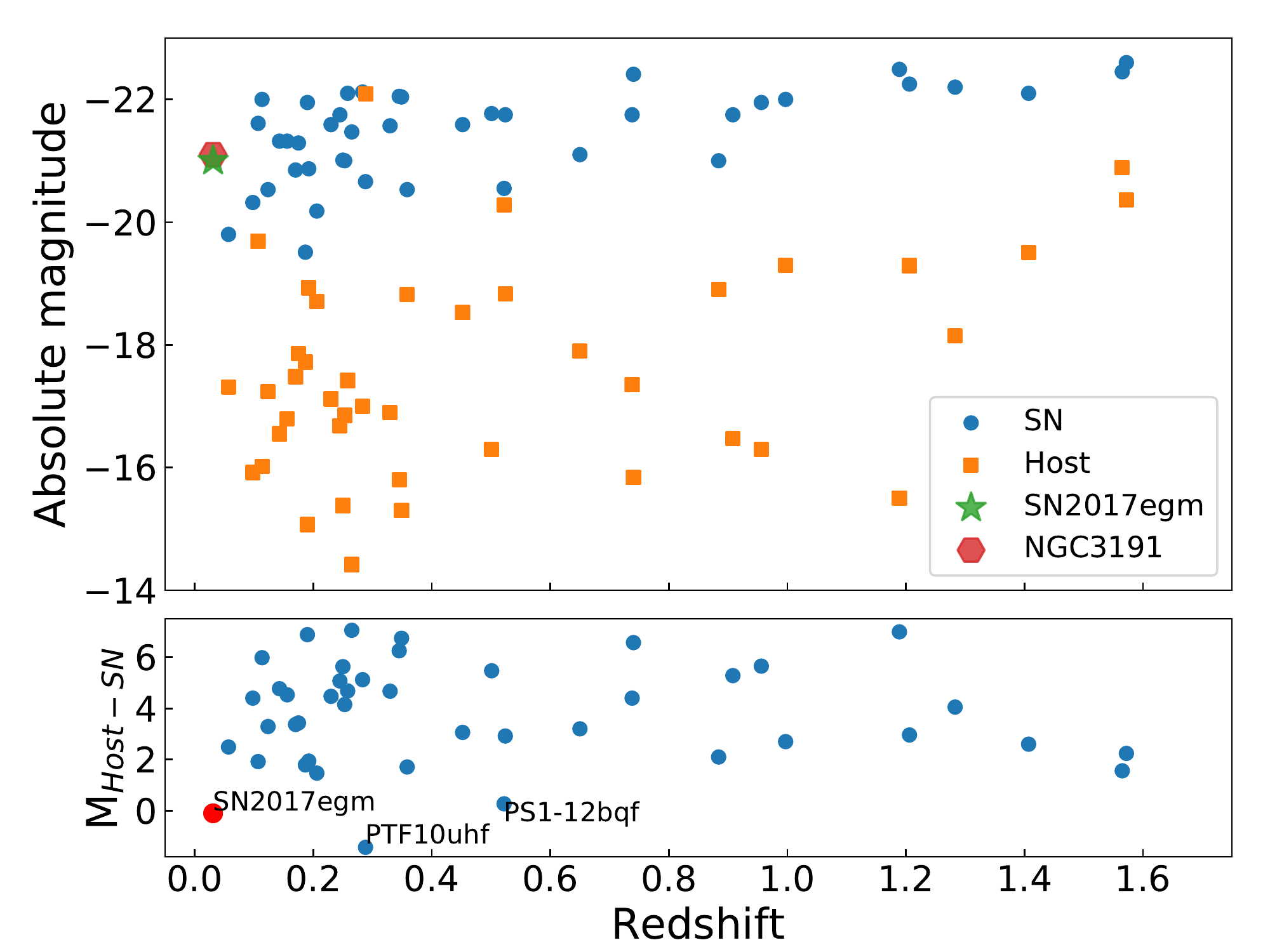}
    \end{subfigure}  
\hfill
    \begin{subfigure}[h]{0.45\linewidth}
        \includegraphics[width=\columnwidth]{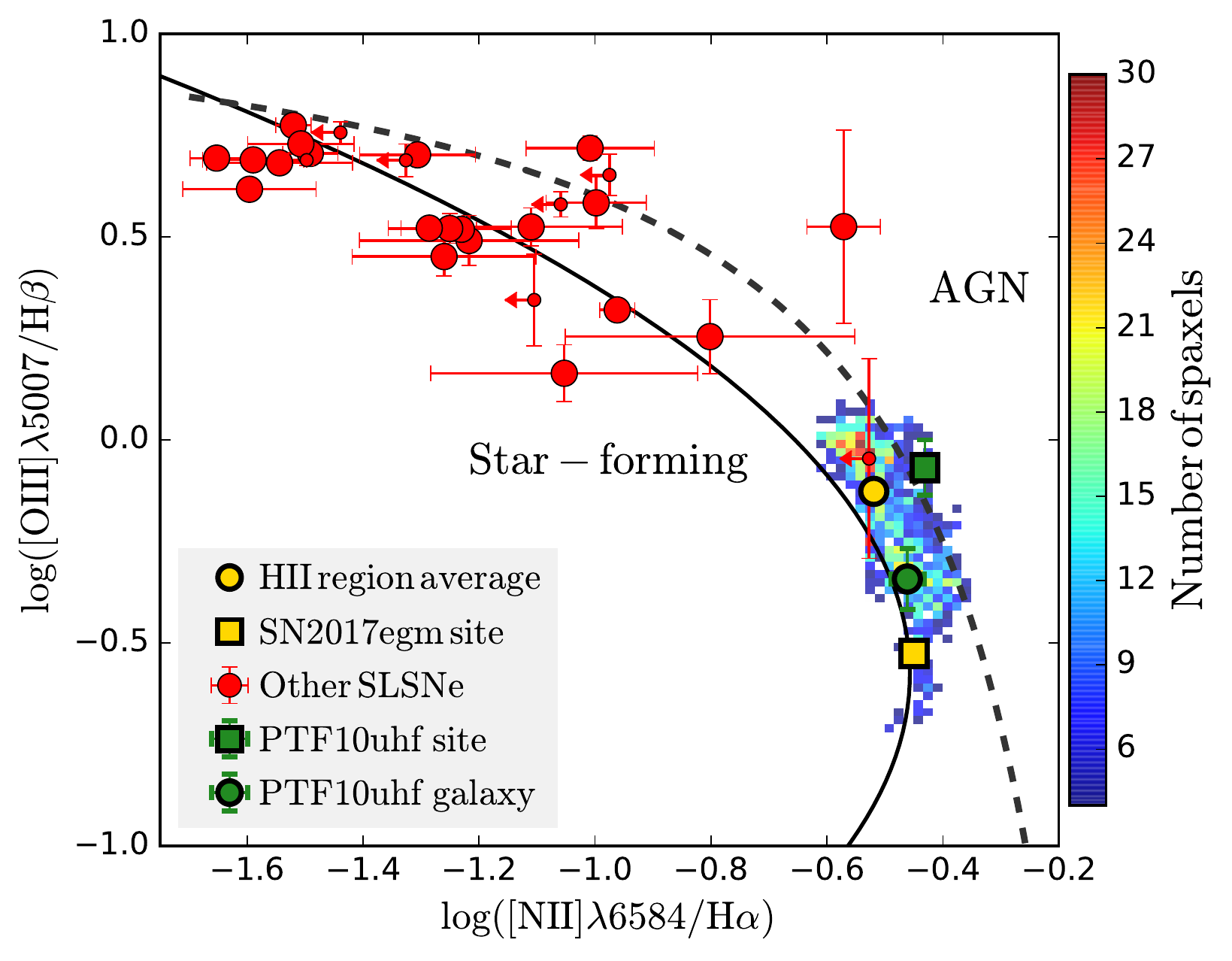}
    \end{subfigure}
\caption{{\it Left:} Redshift evolution of absolute magnitudes of SLSNe and their host galaxies. The upper panel shows $M_{g}$ of SLSNe (blue circle) compared with the host $M_{g}$ (orange square). SN~2017egm and its host NGC~3191 are highlighted. The lower panel shows the magnitude difference between the SLSNe and their host galaxies. {\it Right:} Spaxel BPT diagram for NGC~3191. The dotted line illustrates the indicative separation (at $z=0$) of AGN and star-forming galaxies \citep{2013ApJ...774L..10K}. The solid line expresses the ridge line of SDSS galaxies \citep{2008MNRAS.385..769B}. 
NGC~3191 and the SN locations are similar to the massive host of PTF~10uhf.}
\label{fig_BPT}
\end{figure*}

\subsection{Selection effect}
We calculated a host absolute $g$-band magnitude of $M_{g}=-21.1$ adopting from the SDSS DR14 Petrosian mag. SN~2017egm had a peak $M_{g}=-21$, which is at the fainter end of the SLSN luminosity function \citep[e.g.][]{2014ApJ...796...87I, 2015MNRAS.452.3869N, 2017arXiv170801619L}, and from this we obtain a difference in the host galaxy and SN peak magnitude of $M_{Host-SLSN}\sim0$ mag.

We updated the redshift evolution of the peak absolute magnitudes of SLSNe ($M_{g}$) and their host galaxies (Fig.\,\ref{fig_BPT}), which was first shown by \cite{2015MNRAS.448.1206M} to include data from \cite{2014ApJ...787..138L, 2015MNRAS.452.3869N, 2016ApJ...830...13P, 2016arXiv161205978S, 2017MNRAS.470.3566C, 2017A&A...602A...9C, 2017arXiv170801619L, 2017arXiv170801623D}.
SLSNe are in general 2 to 6 magnitudes brighter than their host galaxies at $z<1$, and the $M_{Host-SLSN}$ is about 1 to 3 magnitudes at higher redshifts. This decrease in the contrast between the host and the SLSN absolute magnitude is consistent with the SLSN host galaxy luminosity evolution to higher luminosities at higher redshifts first reported in \cite{2015MNRAS.448.1206M} and more recently by \citet{2016arXiv161205978S}. 
Only host galaxies of SN~2017egm, PS1-12bqf and PTF~10uhf show a comparable brightness to their SLSNe.
We note that some SLSNe are located in the outskirts of their hosts, hence the difference in magnitudes between them could be even larger if the real host is undetected.

\subsection{BPT diagram}

We plotted the emission line flux ratios [\ion{O}{3}]/H$\beta$ against [\ion{N}{2}]/H$\alpha$ for individual spaxels on the Baldwin-Phillips-Terlevich, or BPT diagram \citep{1981PASP...93....5B}. We found that the spaxels of NGC~3191 were distributed within the star-forming region of the BPT diagram, thus ruling out an active galactic nuclei or shocks as the ionization source, which instead must originate from the radiation from massive stars (Fig.~\ref{fig_BPT}).

Furthermore, we collected emission line fluxes of SLSN host galaxies from the literature \citep{2014ApJ...787..138L, 2015MNRAS.449..917L, 2016ApJ...830...13P, 2017MNRAS.470.3566C, 2017A&A...602A...9C} and corrected for foreground and internal dust extinction given by the Balmer decrement.  
SLSN host galaxies are located predominantly in the high ionization region with $\log$([\ion{O}{3}]/\mbox{H$\beta$}\,)$>0.5$ for any given [\ion{N}{2}]/\mbox{H$\alpha$}. This region is populated by
extreme emission line galaxies as \citealt{2015MNRAS.449..917L} pointed out.
The positions of SN~2017egm and PTF~10uhf are in a distinct region of the BPT diagram to most other SLSN hosts. This may indicate that SLSNe extend out to large metallicities than previously believed, and may cover a similar region to GRB host galaxies \citep[see Fig. 8 in][]{2015A&A...581A.125K}. Improved completeness of low redshift SN samples should indicate if SN~2017egm (and PTF~10uhf) are outliers or at the extreme end of a continuous distribution.

\subsection{Stellar ages and progenitor masses of GRBs and SLSNe}

We used the EW of \mbox{H$\alpha$}\, as a tracer of the stellar population age (e.g. \citealt{2011A&A...530A..95L}; \citealt{2016A&A...593A..78K} and references therein). In general a higher EW$_{H\alpha}$ corresponds to a younger age, though the relation between the EW and age is dependent on the star-formation history, initial-mass function, metallicity, binary or single stellar populations \citep[e.g.][]{2009MNRAS.400.1019E}.

We collected the EWs of \mbox{H$\alpha$}\, at the explosion sites of SLSNe and GRBs from IFU and resolved long-slit spectroscopy (for references see Table\,\ref{tab:comp_ifu}),
and estimated the stellar population ages using consistent models.
We applied the Binary Population and Spectral Synthesis (BPASS\footnote{\texttt{http://bpass.auckland.ac.nz}} version 2.1 models \citep[][Eldridge et al. in prep.]{2016MNRAS.456..485S} in combination with the potoionization Code CLOUDY \citep[e.g.][]{2013RMxAA..49..137F}. These models will be described in Xiao, Stanway \& Eldridge (in prep.), and have been used to study the ages of SNe in \citet{2017arXiv170503606X}. 

To calculate the ages of the stellar populations we select models that have [O/H] value closest to that measured at the explosion site using the O3N2 calibration (Table\,\ref{tab:comp_ifu}). 
For each of the SLSNe we then provide two ages, one assuming the stellar population is comprised of single stars only and the second of binary stars. We see that there is a significant difference 
in the stellar population ages that we derive depending on the BPASS model used. 
The reasons for this difference are first that mass transfer from primary stars to their companions, as well as possible stellar mergers, can form massive stars that are older than if they had formed through single stellar evolutionary channels. Secondly, binary interactions remove the hydrogen envelope from stars that cannot lose their envelopes through stellar winds. These stars are effectively the hot helium cores of more massive stars and are expected to go on and become the progenitors of typical type Ib/c SNe \citep{2013MNRAS.436..774E, 2017arXiv170107439G}.

The ages do not differ greatly between SLSNe and GRBs, although the models imply that GRB progenitors are similar and much younger than SLSNe.
In most cases the binary models predict lower initial masses except in the case of PTF~12dam, where the initial mass is relatively high independent of the stellar population models used (Table\,\ref{tab:comp_ifu}). 

The zero age main sequence (ZAMS) masses predicted from single star models in the case of SN~2017egm and PTF~11hrq are compatible with those inferred from the nebular spectra of slow-evolving SLSNe \citep{2017ApJ...835...13J}.
The initial masses implied from binary models are significantly lower, which would correspond to a SLSN ejecta mass on the order of 0.1-1 M$_\odot$. Such a small ejecta mass is unlikely to produce the observed light curve if it is powered by an internal source. Alternatively the SLSNe may be powered by interaction, although this is not supported by the observed spectral evolution.

For GRB~980425A and GRB ~100316D, which were accompanying by bright type Ic SNe, the low ZAMS progenitor masses predicted by the binary stellar population models are also somewhat controversial since they would not collapse to form a black hole, which is widely accepted to be the central engine of long GRBs. Although magnetar central engines have also been proposed for some GRBs, in most cases it is difficult to extract sufficient energy from a magnetar to power both the GRB and accompanying SN (e.g. \citealt{2016MNRAS.457.2761C} and references there in). On the other hand, GRBs~111005A and GRB~060505 had no accompanying SN down to deep limits, suggestive of a different underlying progenitor to the standard population of long GRBs, and thus the low ZAMS masses derived from the binary stellar population models are more viable.

Nevertheless, an important point to note is that the measured EWs at the transient position for most of the GRBs and SLSNe listed in Table\,\ref{tab:comp_ifu} are in fact averaged over several stellar populations, because the spatial resolution of the data is not sufficiently high to resolve individual star forming regions. The true EW at the transient position could therefore be larger. In the future it will be valuable to reobserve these transient galaxies at higher spatial resolution once the MUSE narrow-field mode is available, which combined with sophisticated stellar synthesis models such as BPASS, will allow an in depth analysis on the progenitor ZAMS masses of long GRBs and SLSNe.

\begin{table*}[ht!]
\caption{Physical properties inferred from IFU and resolved long-slit spectroscopy at SLSN and GRB explosion sites. For consistency, we re-calculate the stellar population age using the BPASS model. References are: (a) this work, (b) \citealt{2017MNRAS.469.4705C}, (c) \citealt{2015MNRAS.451L..65T}, (d) \citealt{2017A&A...602A..85K} (GRB~980425/SN~1998bw), (e) \citealt{2017arXiv170405509I} (GRB~100316D/SN~2010bh), (f) \citealt{2014MNRAS.441.2034T} (GRB~060505/SN-less), (g) \citealt{2017arXiv170806270T} (GRB~111005A/SN-less).} 
\label{tab:comp_ifu}
\centering
\begin{tabular}{l c c c c c c c}
\hline\hline 
Object & SN~2017egm$^{a}$ & PTF~11hrq$^{b}$ & PTF~12dam$^{c}$&
GRB~980425$^{d}$ & GRB~100316D$^{e}$ & GRB~060505$^{f}$ & GRB~111005A$^{g}$\\ 
\hline
redshift & 0.031& 0.057& 0.107 & 0.0087 & 0.059& 0.089 & 0.013\\
$R_{23}$ (KK04) & $9.11\pm0.01$ & - & $8.19\pm0.13$ &- & - & - & -\\
O3N2 (PP04) & $8.77\pm0.01$ & $8.19\pm0.01$ & $8.01\pm0.14$& $8.31\pm0.01$ &$8.21\pm0.02$ & $8.24\pm0.00$ & $8.63\pm0.03$\\
EW (\mbox{H$\alpha$}) [\AA] & $33.7\pm3.3$  & $60.8\pm2.0$& $764\pm10$& $92\pm15$& $202.46\pm17.79$ & $57\pm5.9$ & $16\pm2$ \\
Age (binary) [Myr] & 25-40& 22-25 & 4-6& 12-18& 15-20&25 & 6-8 \\
ZAMS (binary) [$M_{\odot}$] & 8-11&10-11 &64-66 &13-17 & 12-15&10-11 & 6-7 \\
Age (single) [Myr] & 8-10& 8-10& 1-5& 6-8& 6&8-10 & 1-1.3\\
ZAMS (single) [$M_{\odot}$] & 19-24& 20-26&46-100 & 26-33& 34-39&20-26 & 16-20\\
\hline 
\end{tabular}
\end{table*}

\acknowledgments
We acknowledge Thomas~Kr\"uhler for the data analysis pipeline and Anders~Jerkstrand for useful discussion. We thank the referee, Steve~Schulze and Lin~Yan for helpful comments, and Matt~Nicholl for providing SLSN data points.
TWC and PS acknowledge the support through the Sofia Kovalevskaja Award (Alexander von Humboldt Foundation). LX thanks the China Scholarship Council for funding her PhD study at the University of Auckland and the travel funding and support from the University of Auckland. SJS acknowledges funding from the European Research Council Grant agreement n$^{\rm o}$ [291222] and STFC grant ST/P000312/1.

\end{document}